\RequirePackage{fix-cm}

\documentclass[smallextended]{svjour3VISbi}       
\smartqed  
\usepackage{graphicx}
\usepackage{amssymb}
 \journalname{Quantum Studies: Mathematics and Foundations (2017) 1-22}
 \doi{DOI: 10.1007/s40509-017-0119-9}
\newcommand{\Schrodinger}{Schr\"{o}dinger~}
\begin{document}

\bibliographystyle{spphys}       

\title{Hydrodynamics of Superfluid Quantum Space: \\ particle of spin-1/2 in a magnetic field}


\titlerunning{Hydrodynamics of Superfluid Quantum Space: particle of spin-1/2 ...}        

\author{Valeriy I. Sbitnev
}


\institute{V. I. Sbitnev \at
              St. Petersburg B. P. Konstantinov Nuclear Physics Institute, NRC Kurchatov Institute, Gatchina, Leningrad district, 188350, Russia; \\
               Department of Electrical Engineering and Computer Sciences, University of California at Berkeley, Berkeley, CA 94720, USA\\ \\
              Tel.: +781-37-137944\\
              \email{valery.sbitnev@gmail.com}      
}

\date{\today}

\maketitle

\begin{abstract}
The modified Navier-Stokes equation describing the velocity field in the superfluid quantum space is loaded by the external Lorentz force introducing electromagnetic fields. In order to open the path for getting the \Schrodinger-Pauli equation describing the behavior of a particle with spin-1/2 in the magnetic field we need to extend the continuity equation to take into account conservation of  spin flows on the 3D sphere. This extension includes conservation of the density distribution function in 6D space, that is a multiplication of the 3D Euclidean space by the 3D sphere of unit radius. The special unitary group SU(2) underlies the rotations of the spin  on this sphere. This group is isomorphic to the group of quaternions containing the real 4x4 matrices of norm 1. Transition to the quaternion group opens up the way to the possibility of describing the spin-1/2 behavior in a magnetic field  as a motion of a spin flag on the 2D sphere. Maxwell's electromagnetic field theory manifests itself in the quaternion group basis by the natural manner.


\end{abstract}

\section{\label{sec1}Introduction}

This article represents a continuation of the previous work~\cite{Sbitnev2017} entitled "Hydrodynamics of superfluid quantum space: de Broglie interpretation of the quantum mechanics". In this work it had been shown that the modified Navier-Stokes equation together with the continuity equation underlie the inference of the \Schrodinger equation. In addition the first equation can contain solutions of vortices simulating spins of particles. The particle is accompanied by de Broglie pilot-wave that is the solution of the \Schrodinger equation. As for the particle spin this equation says nothing about its behavior. 

To clarify the situation with the spin, we introduce the electromagnetic field. The introduction of the external Lorentz force  in the Navier-Stokes equation causes the appearance of scalar and vector electromagnetic potentials deforming wavy motions in the superfluid quantum space~\cite{SbitnevFedi2017}. It is known that the behavior of the particle spin is extremely sensitive to variations of the magnetic field through which the particle passes. So we pay here close attention to the solution of this problem. As a result of this consideration we conclude that the continuity equation should be extended up to including continuity of spin flows on 3D sphere. So the extended continuity equation is the equation acting on 6D space what is represented by product of two 3D spaces - The 3D Euclidean space of the particle positions and the 3D sphere on which rotation of the particle spin occurs.  For that reason we invoke the Clifford algebra~\cite{Baylis1996,Hestenes2015,Lounesto2001,Todorov2011} which permits to consider motion of spins on the 3D sphere in detail~\cite{AgamalyanEtAl1988,IoffeEtAl1991,Sbitnev1989,Sbitnev2008}.

The article is organized as follows.
In addition to the first section, it contains 5 sections.  Sec.~{\ref{sec2}} begins from introduction of the modified Navier-Stokes equation and loading it by the external Lorentz force. Here we define the irrotational and solenoidal velocities that lead to two equations resulting from the Navier-Stokes equation. The one equation is close to the Hamilton-Jacobi equation. And the other equation is the equation for the vorticity. Sec.~{\ref{sec3} defines the extended continuity equation acting in 6D space represented by the product of two 3D spaces. The first space is the 3D Euclidean space and the other space is the 3D sphere of unit radius. In this section we also define spin rotations and quaternion representations of these rotations. And as an example we consider the neutron spin resonance in periodic magnetic fields. Based on the above equations we get  in Sec.~\ref{sec4}  the \Schrodinger-Pauli equations. Sec.~{\ref{sec5}} gives a natural description of the Maxwell's electromagnetic theory in the quaternion basis. Sec.~{\ref{sec6}} represents the concluding remarks

\section{\label{sec2}The Navier-Stokes equation in the electromagnetic field}
 
The modified Navier-Stokes equation looks as follows
\begin{equation}
 m \biggl(
 {{\partial {\vec {\mathit v}}}\over{\partial\,t}}
 + ({\vec {\mathit v}}\cdot\nabla){\vec {\mathit v}}
       \biggr) 
  =    {\vec f({\vec r},t)}    
   \;-\;
     \nabla Q
 \; +\; m\nu(t)\,\nabla^{\,2}{\vec {\mathit v}} .
\label{eq=1}
\end{equation}
Here $Q=P/\rho$ is the quantum potential ($P$ is a pressure of the Bose gas which constitutes the basis of the superfluid quantum space and $\rho$ is the density distribution of particles of mass $m$), and $\nu(t)$ is the kinetic viscosity coefficient that is a fluctuating function satisfying to the following conditions
\begin{equation}
 \langle \mu(t) \rangle = 0_{+}, ~~~~~
 \langle \mu(t)\mu(0) \rangle >0.
\label{eq=2}
\end{equation} 
These conditions tell us that there is an exchange between the kinetic energy of the particle with the superfluid quantum space (SQS) which manifests itself as the intrinsic fluctuations of vacuum~\cite{Nelson1967,Nelson1985,Nelson2012}. This finds a good agreement with the idea of Vigier~\cite{Vigier1988} according to which vacuum behaves like a stochastic covariant superfluid aether whose excitations can interfere with the propagating particles.

  The external force ${\vec f}({\vec r},t)$ in the modified Navier-Stokes equation~(\ref{eq=1}) consists of a sum of two forces, 
  ${\vec f}_{1}({\vec r},t)$ and ${\vec f}_{2}({\vec r},t)$,  as a minimum. 
  The first force we believe is conservative ${\vec f}_{1}({\vec r},t) = -\nabla U({\vec r},t)$,
   where $U({\vec r},t)$ is the external potential.
  The second force let be the Lorentz force
\begin{equation}
  {\vec f}_{2} = \rho_{q}{\vec E} + {\vec J}_{q}\!\times\!{\vec B}.
\label{eq=3}
\end{equation}
 Here, $\rho_{q}=q/{\Delta V}=eN/{\Delta V}=e\rho$ is the charge $q=eN$ per the unit volume $\Delta V$ ($e$ is the electron charge), 
${\vec J}_{q} = q{\vec{\mathit v}}/{\Delta V}=e\rho{\vec{\mathit v}}$ is the density current, 
${\vec E}$ is an electric field, and ${\vec B}$ is a magnetic field. Both are external fields.

  Further we express the electric field $\vec E$ through the vector, $\vec A$, and scalar, $\phi$, potentials: ${\vec E} = -\nabla\phi-\partial {\vec A}/\partial\,t$ (here, we adopted the SI unit).
Also, we take into account ${\vec J}_q\!\times\!\vec B = -\vec B\!\times\!{\vec J}_q$.
 So, the Lorentz force takes the view
\begin{equation}
  {\vec f}_{2} =
   - e\rho\biggl(
                     {{\partial {\vec {\mathit A}}}\over{\partial\,t}}  + \nabla \phi
             \biggr) 
   - e\rho  [{\vec B}\times{\vec {\mathit v}}]
\label{eq=4}
\end{equation}
 
   Taking into account the above said we rewrite the modified Navier-Stokes equation~(\ref{eq=1}) in the following form
\begin{eqnarray}
\nonumber
&&
m_{}\biggl(
             {{\partial {\vec {\mathit v}}}\over{\partial\,t}}   
             + \underbrace{{{1}\over{2}}\nabla {\mathit v}^{2} + [{\vec\omega}\times{\vec {\mathit v}}]}
         \biggr)
  + e  \biggl(
              {{\partial {\vec {\mathit A}}}\over{\partial\,t}}  + \nabla \phi
                           \biggr) + e  [{\vec B}\times{\vec {\mathit v}}]
\\
&&                           
= -\nabla \bigl(
                   Q({\vec r},t)  +  U({\vec r},t)
            \bigr)
+ m_{}\nu(t) \nabla^{\,2}  {\vec {\mathit v}}.
\label{eq=5}            
\end{eqnarray}
   The term captured by the brace comes from $({\vec {\mathit v}}\cdot\nabla){\vec {\mathit v}}$.
   The first term in the brace is the kinetic energy and
   the vector ${\vec\omega} = [\nabla\times{\vec{\mathit v}}]$ is called vorticity. 

\subsection{\label{sec21}Potential (curlfree) and rotational (divergencefree) vector fields} 
   
   The fundamental theorem of the vector calculus, Helmholtz's theorem, states that any vector field can be expressed through   the sum of irrotational and solenoidal fields~\cite{Lighthill1986}.
   It can be achieved by the Helmholtz-Hodge decomposition~\cite{AbrahamEtAl2001}. The decomposition suggests that any smooth vector field can be uniquely represented by the sum of its potential (curlfree), rotational (divergencefree) and harmonic (both curlfree and divergencefree) components.  The  potential (curlfree) or irrotational velocity is proportional to gradient of a scalar field $S$.  We will label it by subscript $S$, namely, we will write ${\vec{\mathit v}}_{S}$. Along with potential (curlfree) velocity there is also the rotational or solenoidal velocity. The latter we sign by subscript $R$, and write ${\vec{\mathit v}}_{R}$.
The current velocity ${\vec{\mathit v}}$ can be represented as consisting of two components 
$ {\vec{\mathit v}} = {\vec{\mathit v}}_{{S}} + {\vec{\mathit v}}_{{R}}$. 
 These velocities relate to vortex-free and vortex motions of the fluid medium, respectively. 
 They satisfy the conditions~\cite{Segel2007}
\begin{equation}
\label{eq=6}
 \left\{
    \matrix{
           (\nabla\cdot{\vec{\mathit v}}_{{S}}) \ne 0, & [\nabla\times{\vec{\mathit v}}_{{S}}]=0, \cr
           (\nabla\cdot{\vec{\mathit v}}_{{R}})  =  0, &\,\, [\nabla\times{\vec{\mathit v}}_{{R}}]={\vec\omega}. \cr
           }
 \right.
\end{equation}
One can trace a parallelism between these equations and the Maxwell equations for electric and 
magnetic fields~\cite{Martins2012,MartinsPinheiro2009}. 
In this key, we may represent the kinetic momentum 
${\vec{\mathit p}}=m_{}{\vec{\mathit v}}$ and the kinetic energy as follows
\begin{equation}
\label{eq=7}
 \left\{\,
    \matrix{
           {\vec{\mathit p}} = m_{}({\vec{\mathit v}}_{{S}} + {\vec{\mathit v}}_{{R}})
            = \nabla S - e{\vec A},
\cr\cr
           {\displaystyle
            m_{}{{({\mathit v}_{{S}} + {\mathit v}_{{R}})^2}\over{2}}} = 
           {\displaystyle{{1}\over{2m_{}}}}(
                       \nabla S - e{\vec A}
                                        )^2 .
\cr
           }
 \right.
\end{equation}
 Here, $S$ is a scalar function called the {\it action}, 
 the gradient of which determines the irrotational  velocity ${\vec{\mathit v}}_{{S}}=\nabla S/m$, 
whereas the vector potential underlies determination of the rotational velocity, 
namely, ${\vec{\mathit v}}_{{R}}=-(e/m){\vec A}$.

 Taking into account $m_{}{\vec{\mathit v}} =\nabla S - e{\vec A}$ we rewrite Eq.~(\ref{eq=5})  in detail
\begin{eqnarray}
\nonumber 
&&
    {\frac{\partial\;}{\partial\,t}} (\nabla S - e{\vec A}) 
+  {\frac{1}{2m_{}}}\nabla (\nabla S - e{\vec A})^2
\,+\, m [{\vec\omega}\!\times\!{\vec{\mathit v}}]
\\
&+&
   e  {\frac{\partial\;}{\partial\,t}} {\vec A}  \,+\, e\nabla\phi
 \,+\, e[{\vec B}\!\times\!{\vec{\mathit v}}]
- m_{}{\nu}(t)\nabla^{2}{\vec{\mathit v}}
+ \nabla(Q + U) = 0.
\label{eq=8}
\end{eqnarray} 
  First, one can see that both terms $e{\partial {\vec A}}/{\partial t}$ in this expression cancel each other.
Also, the term  $m[{\vec\omega}\!\times\!{\vec{\mathit v}}]$ can be transformed to $-e[{\vec B}\!\times\!{\vec{\mathit v}}]$  as soon as  we apply the operator curl to the kinetic momentum ${\vec{\mathit p}}=m{\vec{\mathit v}}$, see Eq.~(\ref{eq=7}). We get
\begin{equation}
 m{\vec\omega} = m[\nabla\!\times\!{\vec{\mathit v}}_{R}] = -e[\nabla\!\times\!{\vec A}] = -e{\vec B}.
\label{eq=9}
\end{equation}
The vorticity, $\vec\omega$, is antiparellel to the magnetic field $\vec B$.
 From here it follows that contribution of the magnetic field cancels the vorticity in Eq.~(\ref{eq=8}), namely:  
 $e[{\vec B}\!\times\!{\vec{\mathit v}}]+ m [{\vec\omega}\!\times\!{\vec{\mathit v}}]
 =e[{\vec B}\!\times\!{\vec{\mathit v}}] - e [{\vec B}\!\times\!{\vec{\mathit v}}]=0$.
   The Lorentz force applied to the fluid medium avoids the production of vorticity and the consequent onset of the self-maintaining turbulent flows~\cite{FanDong2016}. 
  
  In the light of the above remarks we may rewrite Eq.~(\ref{eq=8}) as follows
\begin{equation}
 \nabla
 \biggr(
      {\frac{\partial\;}{\partial\,t}} S + {\frac{1}{2m_{}}} (\nabla S - e{\vec A})^2 + e\phi
     + (Q  + U)
 \biggl)   
     - {\nu}(t)m\nabla^2 {\vec{\mathit v}}
  = 0.
\label{eq=10}
\end{equation}
Here we write the term $\nu(t)m\nabla^2{\vec{\mathit v}}$ separate from the terms gathered under the big brackets since the velocity ${\vec{\mathit v}}$ is represented by the sum of the irrotational and rotational velocities, ${\vec{\mathit v}}_{S}$ and 
${\vec{\mathit v}}_{R}$, respectively. Therefore, by multiplying this equation by the operator curl we obtain the equation for the vorticity
\begin{equation}
   \nabla^{2} {\vec\omega} = 0.
\label{eq=11}
\end{equation}
stationary in time since both terms $e{{\partial{\vec A}}/{\partial t}}$ in Eq.~(\ref{eq=8}) cancel  each other.
 
On the other hand, the term $\nu(t)m\nabla^2{\vec{\mathit v}}_{\!R}$ introduces a non-linear noise source in the \Schrodinger-Pauli equation. To provide the transition to this equation
we should add to Eq.~(\ref{eq=10}) the continuity equation. 
However, the continuity equation should act now on the extended 6D space  ${\mathcal R}^{3}\otimes{\mathcal S}^{3}$.
Imagine 3D Euclidean space ${\mathcal R}^{3}$ where each point ${\vec r}=(x, y, z)$ is center of a sphere, around which rotations of the unit vector can occur.
Let us define an unit vector the tip of which, having form of a flag instead arrow, rotates about the vector.  Such a vector associated with rotation in Euclidean space is the spinor
\begin{equation}
  |\varphi({\vec r},t)\rangle =
  \biggl(
     \matrix{
         |\varphi_{\uparrow}\rangle \cr
         |\varphi_{\downarrow}\rangle \cr
     }
  \biggr)
  =
  \biggl(
     \matrix{
              \, s_0+{\bf i}s_z  \cr
        {\bf i}(s_x+{\bf i}s_y) \cr
     }
  \biggr).
\label{eq=12}
\end{equation}
Here $|\varphi_{\uparrow}\rangle$ and $|\varphi_{\downarrow}\rangle$ means upward and downward states of the spin, respectively. As for real numbers $s_0$, $s_x$, $s_y$, $s_z$ they will be explained later on at introducing quaternions.
 Here ${\bf i}$  is the imaginary unit $\sqrt{-1}$.

\section{\label{sec3}The continuity equation}

Let us first define a joint amplitude distribution of the translation and rotation flows on the space ${\mathcal R}^{3}\otimes{\mathcal S}^{3}$ that will be written as ${\mathcal R}({\vec r},t)=R({\vec r},t)|\varphi({\vec r},t)\rangle$. Here $R({\vec r},t)$ is the amplitude distribution of the translation flows in the space ${\mathcal R}^{3}$ and $|\varphi({\vec r},t)\rangle$ is the amplitude distribution of spin flows on the sphere ${\mathcal S}^{3}$.
Let a generator ${\mathcal D}({\vec u})$ realizes a small shift of  ${\mathcal R}({\vec r},t)$   on $\delta\tau$:
\begin{eqnarray}
  {\mathcal R}({\vec r},t+{\delta\tau}) =
  {\mathcal D}({\vec u}){\mathcal R}({\vec r},t).
\label{eq=13}
\end{eqnarray}
 The generator ${\mathcal D}({\vec u})$ transforms the amplitude distribution of the spin flows, 
but does not act on $R({\vec r},t)$. Therefore, the shifting transformation reduces to two the continuity equations:
\begin{eqnarray}
\label{eq=14}
 {{d}\over{d\,t}}\rho({\vec r},t) &=& 0,\\
| \varphi({\vec r},t+{\delta\tau})\rangle  
 &=& {\mathcal D}({\vec u}) | \varphi({\vec r},t)\rangle .
\label{eq=15}
\end{eqnarray}
Here $\rho({\vec r},t) = R^2({\vec r},t)$ is the density distribution of the superfluid quantum medium.
Eq.~(\ref{eq=14}) says that there are neither sources nor sinks in this medium, 
while Eq.~(\ref{eq=15}) implies  existence of external electromagnetic fields that perturb motions of spins.
The generator  ${\mathcal D}({\vec u})$ is as follows~\cite{AgamalyanEtAl1988}
\begin{equation} 
 {\mathcal D}({\vec u}) 
 =         u_{0}\sigma_{0}
 + {\bf i}u_{x}\sigma_{x} 
 + {\bf i}u_{y}\sigma_{y} 
 + {\bf i}u_{z}\sigma_{z} .
\label{eq=16}
\end{equation}
  Here ${\bf i}$  is the imaginary unit. Four basic matrices in this expression, namely,
 three Pauli matrices, $\sigma_{x}$, $\sigma_{y}$, and $\sigma_{z}$, and the unit matrix $\sigma_{0}$ read:
\begin{equation}
  \sigma_{x} = 
  \left(
        \matrix{
                     0  &  1 \cr
                     1  &  0 \cr
                  }
  \right),~
  \sigma_{y} = 
  \left(
        \matrix{
                     0         &   -{\bf i} \cr
                     {\bf i}  &    ~~0    \cr
                  }
  \right),~
   \sigma_{z} = 
  \left(
        \matrix{
                     1  &  ~~0 \cr
                     0  &  -1    \cr
                  }
  \right),~
  \sigma_{0} = 
  \left(
        \matrix{
                     1  &  0 \cr
                     0  &  1 \cr
                  }
  \right).
\label{eq=17} 
\end{equation}
Real coefficients, $u_{0}$, $u_{x}$, $u_{y}$, $u_{z}$, in ${\mathcal D}({\vec u})$, are as follows~\cite{Sbitnev1989}
\begin{equation}
 u_{0} = \cos\biggl({{\xi}\over{2}}\biggr),~
 u_{x} = b_{x}\sin\biggl({{\xi}\over{2}}\biggr),~
 u_{y} = b_{y}\sin\biggl({{\xi}\over{2}}\biggr),~
 u_{z} = b_{z}\sin\biggl({{\xi}\over{2}}\biggr).
\label{eq=18}
\end{equation}
 Here ${\vec b}=(b_{x}, b_{y}, b_{z})$  is the unit vector
\begin{equation}
  {\vec b} = {{\vec B}\over{\sqrt{B_{x}^2+B_{y}^2+B_{z}^2}}}
\label{eq=19}
\end{equation} 
 pointing direction of the magnetic induction ${\vec B}$,
 while $\xi$ is the precession angle proportional to the proper time of the spin in a magnetic field
\begin{equation}
  \xi = -\gamma_{e}\sqrt{B_{x}^2+B_{y}^2+B_{z}^2}\cdot{\delta\tau}.
\label{eq=20}
\end{equation} 
Here $\gamma_{e}=g_{e}\mu_{B}/\hbar$ is the gyromagnetic ratio of electron  (the ratio of the magnetic moment of electron to its angular momentum), $\mu_{B} = e\hbar/2m_{e}$ is the Bohr magneton defined in SI units , $m_{e}$ is the electron mass, and $\hbar$ is the reduced Planck constant. The gyromagnetic ratio for the self-spinning electron is twice bigger than the value for an orbiting electron, therefore the factor $g_{e}$ is about 2.

The above transformation associates the rotation of the sphere ${\mathcal S}^{3}$ of unit radius with shiftings on a complex vector space with using the unitary unimodular group SU(2)~\cite{Hamermesh1962,PenroseRindler1984}.
However, for a more visual imagination, it makes sense to go to  the quaternion representation of the spin rotations~\cite{AgamalyanEtAl1988}.

\subsection{\label{sec31}Quaternion representation}

The generator~(\ref{eq=16})  is element of the group SU(2). Let ${\mathcal D}({\vec u})$ and ${\mathcal D}({\vec s})$ belong to SU(2), then their multiplication 
\begin{eqnarray}
\label{eq=21}
  &&  {\mathcal D}({\vec {\tilde s}}) =  {\mathcal D}({\vec u})\cdot{\mathcal D}({\vec s}) \;\Rightarrow\;
 ({\tilde s}_0\sigma_0 + {\bf i}{\tilde s}_x\sigma_x + {\bf i}{\tilde s}_y\sigma_y + {\bf i}{\tilde s}_z\sigma_z)
\\ \nonumber
&=& 
({u}_0\sigma_0 + {\bf i}{u}_x\sigma_x + {\bf i}{u}_y\sigma_y + {\bf i}{u}_z\sigma_z)\cdot
({s}_0\sigma_0 + {\bf i}{s}_x\sigma_x + {\bf i}{s}_y\sigma_y + {\bf i}{s}_z\sigma_z)
\end{eqnarray}
belongs to the group SU(2) as well.
By putting into account the multiplication order of the Pauli matrices
\begin{eqnarray}
\nonumber
 &&
 \sigma_{x}\sigma_{y}=-\sigma_{y}\sigma_{x}={\bf i}\sigma_{z},~~~~~
 \sigma_{y}\sigma_{z}=-\sigma_{z}\sigma_{y}={\bf i}\sigma_{x},      
 \\
 &&
 \sigma_{z}\sigma_{x}=-\sigma_{x}\sigma_{z}={\bf i}\sigma_{y},~~~~~ 
 \sigma_{x}^{2} = \sigma_{y}^{2} = \sigma_{z}^{2} = \sigma_{0}
\label{eq=22}
\end{eqnarray}   
we find representation of Eq.~(\ref{eq=21}) through 4$\times$4 matrix multiplication
\begin{equation}
\left(
       \matrix{
                  {\tilde s}_{0} \cr
                  {\tilde s}_{x} \cr
                  {\tilde s}_{y} \cr
                  {\tilde s}_{z} \cr
                 }
\right) =
\left(
      \matrix{
                 u_{0}   &   -u_{x}  &   -u_{y}  &   -u_{z}  \cr
                 u_{x}   &~~u_{0}  &~~u_{z}  &   -u_{y}  \cr
                 u_{y}   &   -u_{z}  &~~u_{0}  &~~u_{x}  \cr
                 u_{z}   &~~u_{y}  &   -u_{x}  &~~u_{0}  \cr
                }       
\right)
\left(
       \matrix{
                  s_{0} \cr
                  s_{x} \cr
                  s_{y} \cr
                  s_{z} \cr
                 }
\right)
\label{eq=23}
\end{equation} 

One may extract from Eq.~(\ref{eq=23}) a set of four quaternion matrices~\cite{AgamalyanEtAl1988,Sbitnev2008}:
\begin{eqnarray}
\nonumber
  &&
  \eta_{x} =
  \left(
        \matrix{
                    ~~0 &    -1 & ~~0 & ~~0 \cr
                    ~~1 & ~~0 & ~~0 & ~~0 \cr
                    ~~0 & ~~0 & ~~0 & ~~1 \cr
                    ~~0 & ~~0 &    -1 & ~~0 \cr
                   }
  \right),~
  \eta_{y} =
  \left(
        \matrix{
                    ~~0 & ~~0 &    -1 & ~~0 \cr
                    ~~0 & ~~0 & ~~0 &    -1 \cr
                    ~~1 & ~~0 & ~~0 & ~~0 \cr
                    ~~0 & ~~1 & ~~0 & ~~0 \cr
                   }
  \right),~  \\
  &&
  \eta_{z} =
  \left(
        \matrix{
                    ~~0 & ~~0 & ~~0 &    -1 \cr
                    ~~0 & ~~0 & ~~1 & ~~0 \cr
                    ~~0 &    -1 & ~~0 & ~~0 \cr
                    ~~1 & ~~0 & ~~0 & ~~0 \cr
                   }
  \right),~
  \eta_{0} =
  \left(
        \matrix{
                    ~~1 & ~~0 & ~~~0 & ~~0 \cr
                    ~~0 & ~~1 & ~~~0 & ~~0 \cr
                    ~~0 & ~~0 & ~~~1 & ~~0 \cr
                    ~~0 & ~~0 & ~~~0 & ~~1 \cr
                   }
  \right),~  
\label{eq=24}  
\end{eqnarray}
\begin{eqnarray}
\nonumber
 &&
 \eta_{x}\eta_{y}=-\eta_{y}\eta_{x}=-\eta_{z},~~~~~
 \eta_{y}\eta_{z}=-\eta_{z}\eta_{y}=-\eta_{x},      
 \\
 &&
 \eta_{z}\eta_{x}=-\eta_{x}\eta_{z}=-\eta_{y},~~~~~ 
 \eta_{x}^{2} = \eta_{y}^{2} = \eta_{z}^{2} = -\eta_{0}.
\label{eq=25}
\end{eqnarray}   
One can rewrite the generator~(\ref{eq=16}) in the basis of the quaternion matrices
\begin{equation} 
 {\mathcal D}({\vec u}) 
 = u_{0}\eta_{0}
 + u_{x}\eta_{x} 
 + u_{y}\eta_{y} 
 + u_{z}\eta_{z} .
\label{eq=26}
\end{equation}
 And by rewriting Eq.~(\ref{eq=15}) we get the following equation
\begin{equation}
 |\varphi({\vec r},t+{\delta\tau})\rangle  =  
 ( u_{0}\eta_{0} + u_{x}\eta_{x}  + u_{y}\eta_{y}  + u_{z}\eta_{z})|\varphi({\vec r},t)\rangle.
\label{eq=27}
\end{equation} 
 Now the spinor $|\varphi({\vec r},t)\rangle$ is represented by the column of four real variables $s_0$, $s_x$, $s_y$, and $s_z$.
 The variables  $s_x$, $s_y$, and $s_z$ mark the tip of the vector on the sphere ${\mathcal S}^3$, whereas the variable $s_0$ shows orientation of the flag of this vector. Let us show it on an example of the resonance flip of the spin~1/2 in the periodic magnetic field~\cite{AgamalyanEtAl1988}.

\subsection{\label{sec32}Neutron spin resonance in a periodic magnetic field}

Let be given the periodic magnetic structure shown in Fig.~\ref{fig=1}. 
Here we consider the spin resonance of neutrons flying through the periodic magnetic structure (PMS). Since neutron has no electric charge, we can consider an interaction of the neutron only with the magnetic field. Neutrons flying out of a reactor have different velocities and different orientations of the spins. Therefore, first of all we need to select neutrons with given orientation of their spins~\cite{AgamalyanEtAl1988}. For this aim before PMS the Fe-Co mirror polarizer is placed. Depending on the spin orientation the polarizer reflects the flying neutrons on different angles, as shown by dotted arrows in this figure. The neutrons having the spin orientation, say along $z$ axis, are directed to PMS.  PMS  realizes a revolution of the spins depending on neutrons speed through the PMS. Next the spins with given orientation are selected by the analyzer.
\begin{figure}[htb!]
  \begin{picture}(180,140)(0,5)
      \includegraphics[scale=0.5]{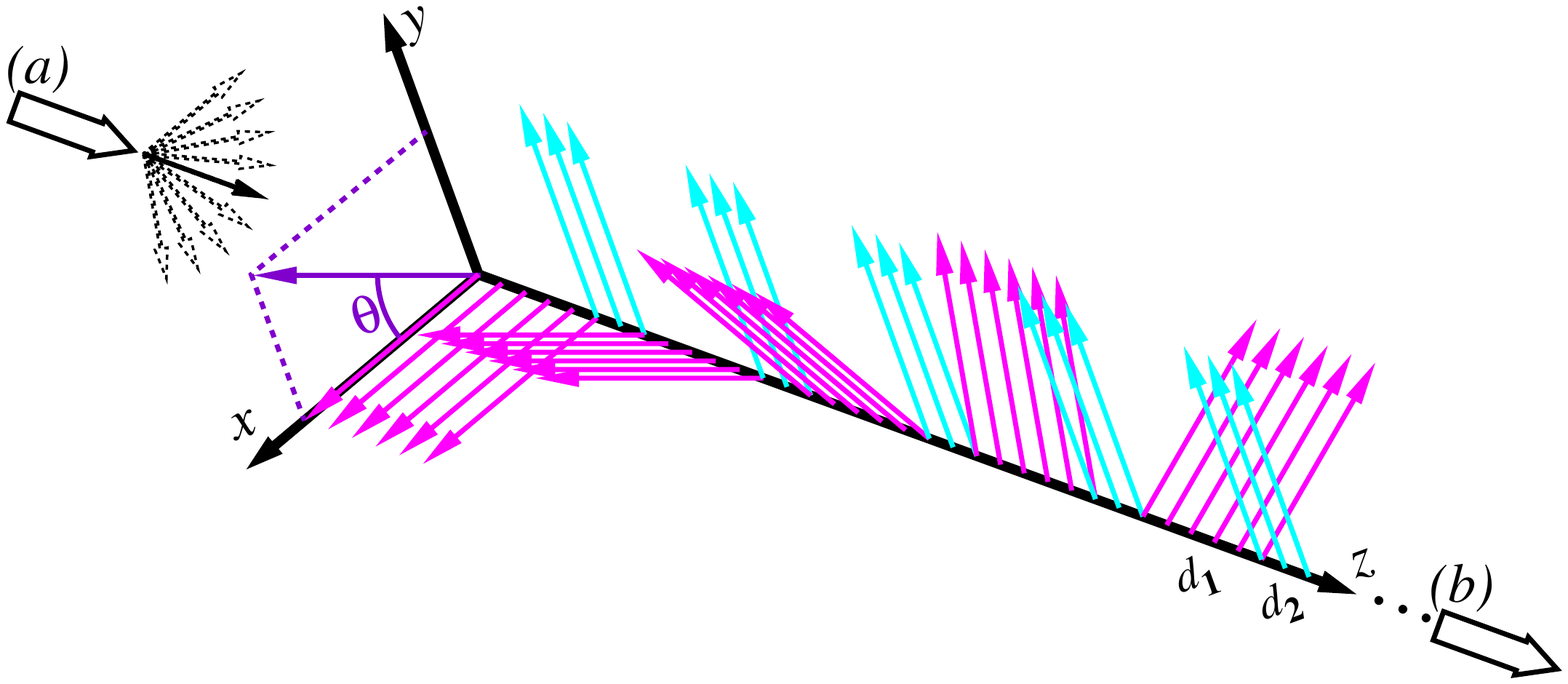}
  \end{picture}
  \caption{
Spatially periodic magnetic structure. Neutrons are flying along $z$ axis. Arrows {\it a} and {\it b} depict conditionally mirror polarizer and analyzer. On the output of the polarizer neutrons have spins oriented along $z$ axis, regardless of their velocities.
The  analyzer permits to select the spins with the orientation that has to be taken into account.
  }
  \label{fig=1}
\end{figure}

PMS is assembled by alternating magnetic bars with thickness of $d_1$, between which  magnetized films with thickness $d_2$ are laid ($d_2\ll d_1$).  Magnetic fields in the films are oriented in parallel each other. Whereas magnetic fields of the bars can be turned out relative to each other by a fixed angle. The coefficients $u_0$, $u_x$, $u_y$, $u_z$ for this configuration have the following view~\cite{AgamalyanEtAl1988}:
\begin{eqnarray}
\label{eq=28}
u_{0,1} &=& \cos\biggl(
                  {{\xi_1}\over{2}}
                 \biggr),\:
 u_{x,1} = \sin\biggl(
                  {{\xi_1}\over{2}}
                 \biggr)\cos(\theta),\:
 u_{y,1} = \sin\biggl(
                  {{\xi_1}\over{2}}
                 \biggr)\sin(\theta),\:
                  u_{z,1} = 0,~~~~~
 \\
  u_{0,2} &=& \cos\biggl(
                  {{\xi_2}\over{2}}
                 \biggr),\:
  u_{x,2} = 0,\hskip64pt
   u_{y,2} = \sin\biggl(
                  {{\xi_2}\over{2}}
                 \biggr),\hskip28pt
      u_{z,2} = 0.~~~~~    
\label{eq=29}              
\end{eqnarray}
 Phase shifts $\xi_1$ and $\xi_2$ result from the formulas
\begin{equation}
\xi_1 = \gamma_n\sqrt{B_x^2+B_y^2}\cdot d_1/{\mathit v},~~~~
\xi_2 =\gamma_n  B_y d_2/{\mathit v}.
\label{eq=30}
\end{equation} 
Here $\gamma_n=2\mu_n/\hbar$ is the neutron's gyro-magnetic ratio, $\mu_n=-1.91e\hbar/2m_nc$ is the neutron's magnetic moment, $m_n$ is its mass. The speed of a neutron passing through a magnetic field {\it B} is ${\mathit v}$.

We have no possibility to show shifts of the coefficients $u_0$, $u_x$, $u_y$, $u_z$ on the 3D sphere. However a projection of these motions on the 2D sphere exists. We will show this projection in sec.~{\ref{sec5}}, when we will deal with Maxwell's electromagnetic theory. In fact, the projection makes a transition from the spin coordinates $(s_0, s_x, s_y, s_z)$ moving on ${\mathcal S}^3$ to the motion of the polarization vector $(P_x, P_y, P_z)$ on ${\mathcal S}^2$. In this case the rotation matrix looks as
\begin{eqnarray}
\nonumber
&& R(u_0, u_x, u_y, u_z) = 2u_0
   \left(
      \matrix{
            ~0    & ~~u_z  &     -u_y \cr
            -u_z  &   ~0     &  ~~u_x \cr
         ~~u_y  &    -u_x  &    ~0    \cr
                 } 
   \right) 
\\ 
   && + 
  \left(
     \matrix{
     1-2(u_y^2+u_z^2) & 2u_xu_y                & 2u_zu_x                  \cr
     2u_yu_x                & 1-2(u_z^2+u_x^2) & 2u_yu_z                  \cr
     2u_zu_x                & 2u_xu_y                 & 1-2(u_x^2+u_y^2)  \cr
               }
  \right) .
\label{eq=31}  
\end{eqnarray}
It should be noted that the polarization vector $(P_x, P_y, P_z)$ contains extra parameter - the coefficient $u_0$, which endows the vector by an additional degree of freedom that can be interpreted as rotation of the flag around this vector.
It is interesting to remark that the flag rotation discloses many common with the topological phase shift~\cite{Berry1984} when someone measures interference effects under a closed cycle of the spin motion on the ${\mathcal S}^3$ sphere~\cite{IoffeEtAl1991}.

Further for the sake of brevity  the matrices $R(u_0, u_x, u_y, u_z)$ will be written simply with subscripts, $R_{a}$, distinguishing the different matrices by the subscripts. Let matrices $R_a$ and $R_b$ belong to the special orthogonal group SO(3), then their product 
$R_a\!\cdot\!R_b$ is a matrix $R_c$ belonging to SO(3) as well. Note that the product $R_b\!\cdot\!R_a$ gives a matrix $R_d \in {\rm SO(3)}$ that is not equal to the matrix $R_c=R_a\!\cdot\!R_b$.

Let the matrix $R_1$ describe rotation of the polarization vector when the neutron passes through the magnetic plate drawn in Fig.~\ref{fig=1} in pink. Its input parameters have been written down in~(\ref{eq=28}). And the matrix $R_2$ describes thr rotation when the neutron passes through the thin film drawn in this figure in cyan. Its input parameters have been written down in~(\ref{eq=29}).
Let us calculate the following stepwise chain
\begin{eqnarray}
\label{eq=32}
  {\vec P}_n &=& (R_2\cdot R_1)^n {\vec P}_0, \\
  {\vec P}_{n,\xi_1} &=& R_1\cdot  (R_2\cdot R_1)^n {\vec P}_0,
\label{eq=33}
\end{eqnarray}
 where ${\vec P}_0 = (0, 0, 1)$ is the initially prepared polarization vector by the mirror polarizer and $n=0,1,2, \cdots, N$, $N$  is amount of the magnetic blocks in PMS (the block is represented by the pair of magnetic plate and the film).                                                                                                                                                                                                                                                                                                     A pair $({\vec P}_n,{\vec P}_{n,\xi_1})$ gives orientation of the arrow on the sphere ${\mathcal S}^2$, the beginning of which is in ${\vec P}_n$ and its end is in ${\vec P}_{n,\xi_1}$. So, a sequence 
\begin{equation}
({\vec P}_0,{\vec P}_{0,\xi_1}), ({\vec P}_1,{\vec P}_{1,\xi_1}), 
 \cdots, ({\vec P}_n,{\vec P}_{n,\xi_1}),  \cdots, 
({\vec P}_N,{\vec P}_{N,\xi_1})
\label{eq=34}
\end{equation}
 shows a path on the sphere ${\mathcal S}^2$ traversed by the tip of the polarization vector when the neutron elapses through PMS containing $N$ magnetic blocks. For small $N$, say $N = 21$, we have a rough partitioning of the magnetic structure $\theta = \pi/N\approx0.15$, $\xi_1 = 0.3$, $\xi_2 = 0.01$.  In this case we observe stepwise shifts of the the polarization vector as shown in Fig.~\ref{fig=2}.  One can see that this stepwise path repeats roughly the helical ring when $2N\cdot\theta = 2\pi$. This case corresponds to the spin resonance.
\begin{figure}[htb!]
  \begin{picture}(180,145)(0,5)
      \includegraphics[scale=0.25]{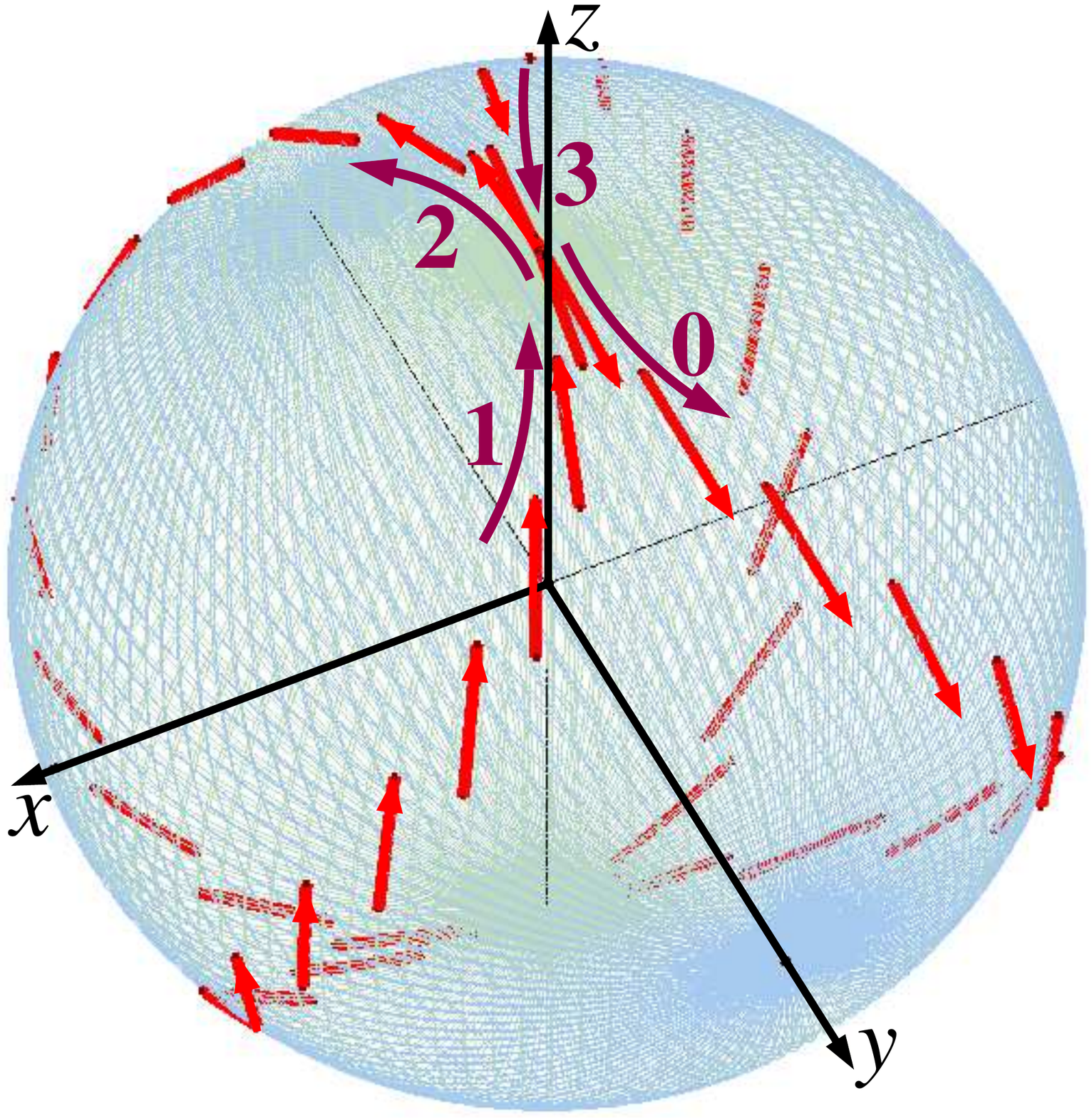}
  \end{picture}
  \caption{Coarse-grained shifts of the the polarization vector ${\vec P}=(P_x, P_y, P_z)$ from its state up, $(0, 0, 1)$,
   along paths 0 to 1, 1 to 2, 2 to 3, 3 to 0: red arrow acquires opposite orientation after revolution on $2\pi$ and gets the same orientation after revolution on $4\pi$.  The parameters for red helical ring are $\xi_1=0.3$, $\xi_2=0.01$, and $\theta=\pi/N\approx0.15$ $N=21$.
  }
  \label{fig=2}
\end{figure}
\begin{figure}[htb!]
  \begin{picture}(180,130)(0,0)
      \includegraphics[scale=0.252]{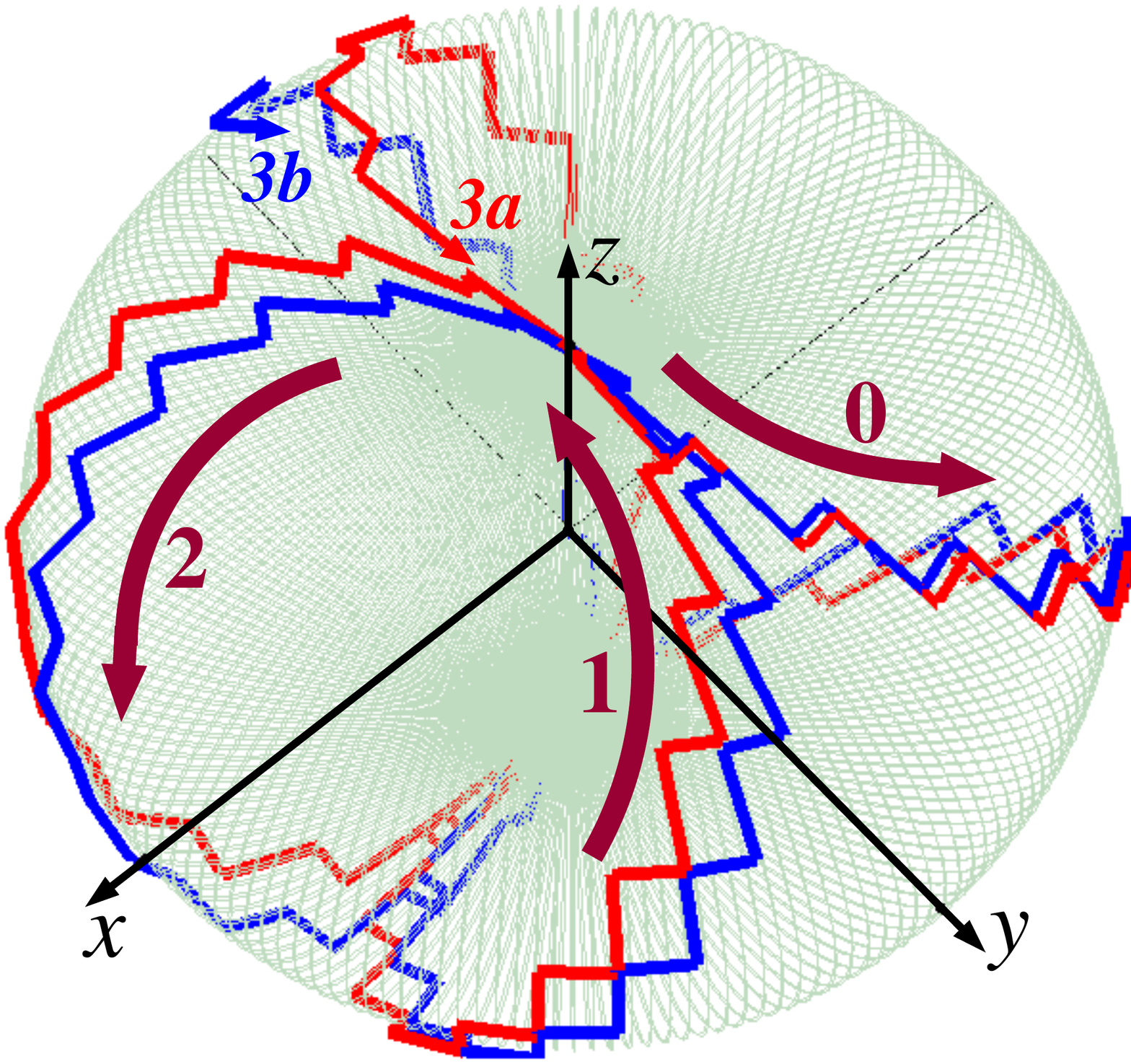}
  \end{picture}
  \caption{Impaired resonance outcome because of  the rapid passage of neutrons through the magnetic structure: red stepwise curve is drawn for the parameters  $\xi_1 =  0.3\cdot0.95=0.285$ and $\xi_2=0.01\cdot0.95=0.0095$  and blue  stepwise curve is drawn for the parameters  $\xi_1 =  0.3\cdot0.9=0.27$ and $\xi_2=0.009$. For $2N\theta=2\pi$ the curves give no closed circles.     
  }
  \label{fig=3}
\end{figure}
 
The helical ring is destroyed when the resonance condition does not occur. For example, let the neutron traveling speed be larger than for the resonant case. As follows from~(\ref{eq=30}) $\xi_1$ and $\xi_2$ will be smaller then 0.3 and 0.01 adopted for the resonance to be considered. Let they be $\xi_1=0.3\cdot0.95=0.285$ and $\xi_2=0.0095$ in the one case and $\xi_1=0.3\cdot0.9=0.27$ and $\xi_2=0.009$ in the other case. These both cases are shown in Fig.~\ref{fig=3} by red and blue colors, respectively. As neutrons complete their passage through the magnetic structure $2N\theta=2\pi$ both  stepwise curves  give no closed circles.   
The last arrows, {\it 3a} and {\it 3b} in Fig.~\ref{fig=3}, do not end in the top pole of the sphere ${\mathcal S}^2$.

 So, emergence of the helical ring results from the spin flip-flap resonance.  However, in order that the helical ring is a smooth continuous ring we must set to zero the parameters of the spatial periodic magnetic structure, namely, $\xi_1\rightarrow0$, $\xi_2\rightarrow0$, and $\theta=\pi/N\rightarrow0$. The last limit means $N\rightarrow\infty$. That is, the thickness of both the magnetic bars and the films tends to zero, and their amount tends to infinity.    
\begin{figure}[htb!]
  \begin{picture}(180,140)(0,5)
      \includegraphics[scale=0.246]{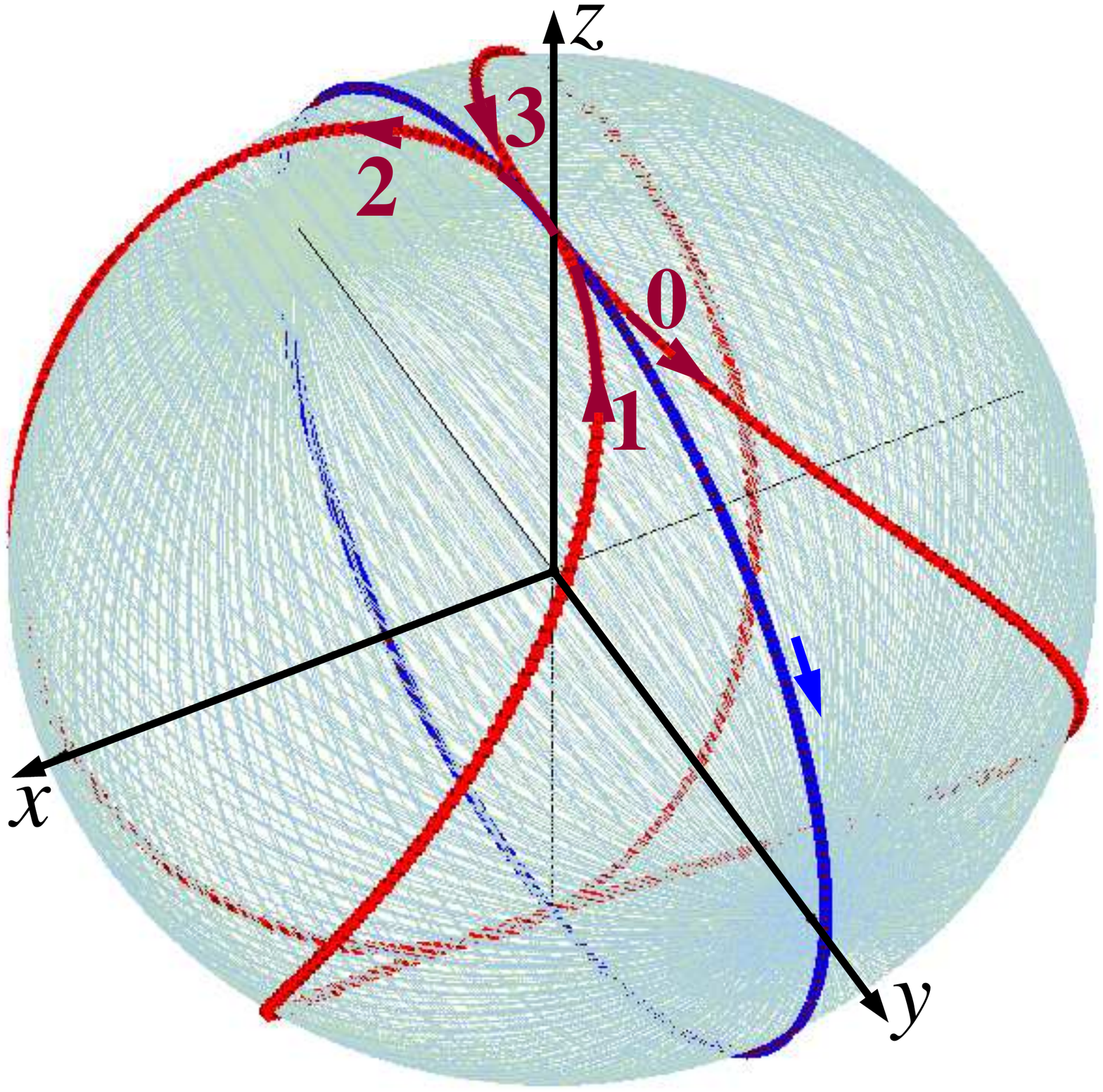}
  \end{picture}
  \caption{Fine-grained shifts of the polarization vector ${\vec P}=(P_x, P_y, P_z)$ from its state up, $(0, 0, 1)$,  along paths 0 to 1, 1 to 2, 2 to 3, 3 to 0: the path acquires opposite direction after revolution on $2\pi$ and gets the initial direction after a revolution on $4\pi$. The parameters for red helical ring are $\xi_1=0.03$, $\xi_2=0.001$, and $\theta=\pi/N\approx0.015$. For blue circle $\theta=0$.
  }
  \label{fig=4}
\end{figure}
\begin{figure}[htb!]
  \begin{picture}(180,130)(0,5)
      \includegraphics[scale=0.25]{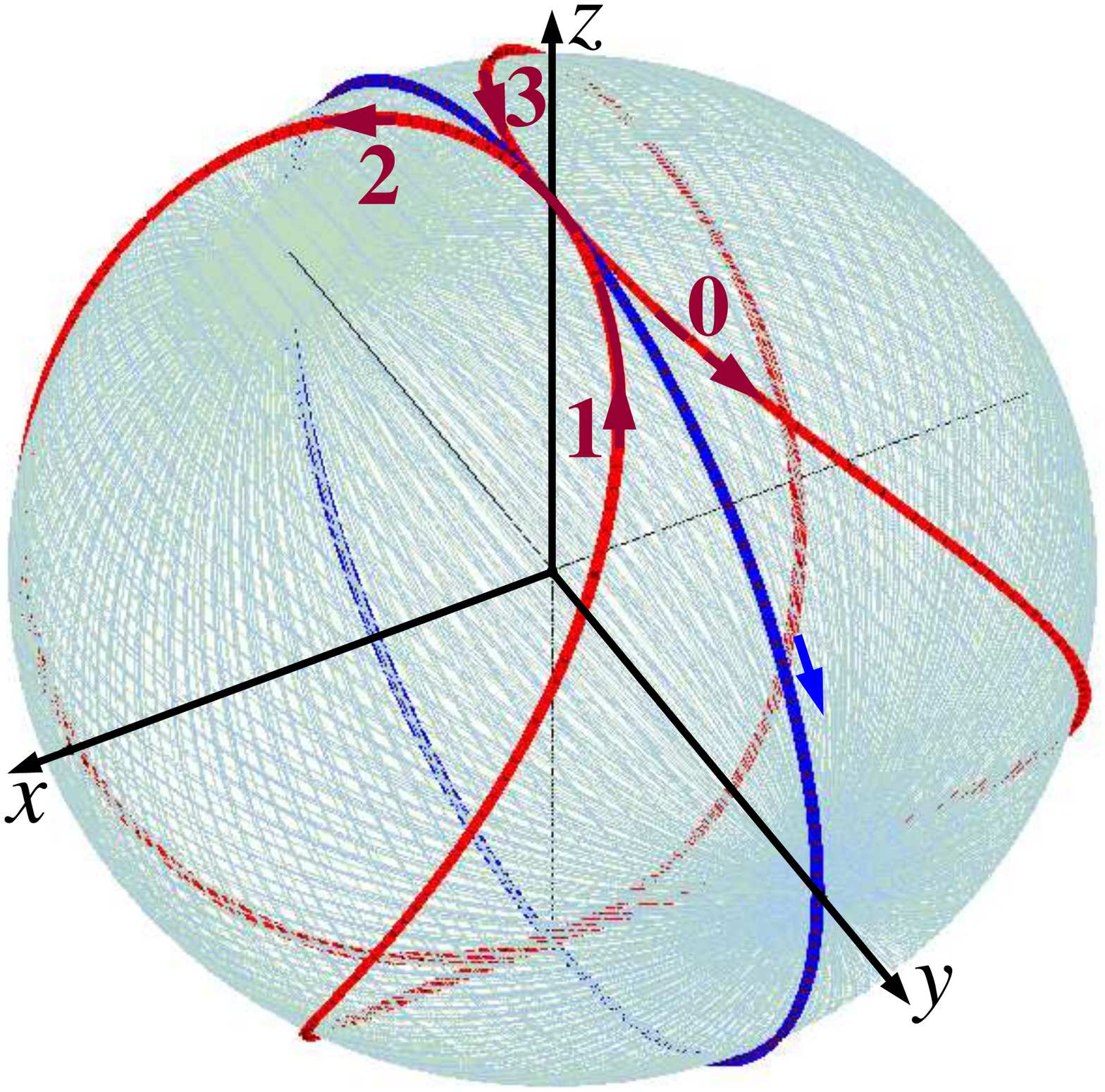}
  \end{picture}
  \caption{Solutions~(\ref{eq=43}), red helical ring, and~(\ref{eq=44}), blue circle, calculated for the parameters ${\mit\Gamma}=0.04$, $\omega=\pi/157\approx0.02$, ${\mit\Delta=0}$.     
  }
  \label{fig=5}
\end{figure}
Fig.~\ref{fig=4} shows emergence of the helical ring in the case of the spin flip-flap resonance for neutrons  passing through periodic magnetic structure having the following parameters $\xi_1=0.03$, $\xi_2=0.001$, $\theta=\pi/N\approx0.015$ at $N=210$. The partition of this structure is on the order more smaller than the previous structure. A small stepwise increment of the red curve can be yet observed. However, as soon as these parameters tends to zero the stepwise increment vanishes.
 
 Blue circle in this figure represents a degenerate case $\theta=0$ - the magnetic fields in all plates have equal orientation lying in $(x,z)$ plane. The magnetic fields in the films lie in $(y,z)$ plane.  They induce weak magnetic perturbations of polarization vector rotation when neutrons pass through the films. These perturbations introduce a weak shift of the polarization vector about $z$ axis. It looks as a weak deviation of the blue circle from the  $(y, z)$ plane.

At ${\delta\tau}$ tending to zero in Eq.~(\ref{eq=27}) we obtain the differential equation describing the spin behavior in a magnetic field
\begin{equation}
  {{d}\over{d\,t}}|\varphi(t)\rangle = -{{\gamma_{n}}\over{2}}({\overrightarrow\eta}\cdot{\vec B}) |\varphi(t)\rangle.
\label{eq=35}
\end{equation} 
 The scalar product $({\overrightarrow\eta}\cdot{\vec B})$ looks as follows
\begin{equation}
 ({\overrightarrow\eta}\cdot{\vec B}) =
   \left(
     \matrix{
                  0    &    -B_x &    -B_y &   -B_z  \cr
                  B_x &   ~0    & ~~B_z &   -B_y  \cr
                  B_y &    -B_z &   ~0    & ~~B_x \cr
                  B_z & ~~B_y &    -B_x &   ~0    \cr
               }
   \right).
\label{eq=36}
\end{equation} 
 The wave 4-vector $|\varphi(t)\rangle$ in the quaternion representation looks as
\begin{equation}
       |\varphi(t)\rangle =
   \left(
     \matrix{
                 s_0(t) \cr
                 s_x(t) \cr
                 s_y(t) \cr
                 s_z(t) \cr
               }
   \right).
\label{eq=37}
\end{equation} 
 As for the spinor, two-component representation, it is shown in Eq.~(\ref{eq=12}).
 
Let us consider a helical magnetic configuration~\cite{IoffeEtAl1991}
\begin{equation}
 {\vec B} = \{
  b\cos(\omega t), b\sin(\omega t), B_z = {\rm const}
                 \}
\label{eq=38}
\end{equation}
Solutions of Eq.~(\ref{eq=35}) with the magnetic field given by~(\ref{eq=38}) look as
\begin{equation}
 \left\{
         \matrix{
 \displaystyle
 s_x(t) = {{\mit\Gamma}\over{\sqrt{\mit\Gamma^2+\mit\Delta^2}}}
 \sin\biggl(
      {{1}\over{2}}t \sqrt{\mit\Gamma^2+\mit\Delta^2}
      \biggr)
 \cos\biggl(
        {{1}\over{2}}\omega t
       \biggr) ,       
\cr
\displaystyle
 s_y(t) = {{-\mit\Gamma}\over{\sqrt{\mit\Gamma^2+\mit\Delta^2}}}
 \sin\biggl(
      {{1}\over{2}}t \sqrt{\mit\Gamma^2+\mit\Delta^2}
      \biggr)
 \sin\biggl(
        {{1}\over{2}}\omega t
       \biggr)   ,     
 \cr
 \displaystyle
 s_z(t) = {{\mit\Delta}\over{\sqrt{\mit\Gamma^2+\mit\Delta^2}}}
 \sin\biggl(
      {{1}\over{2}}t \sqrt{\mit\Gamma^2+\mit\Delta^2}
      \biggr)
 \cos\biggl(
        {{1}\over{2}}\omega t
       \biggr)        
\cr 
\displaystyle
 -   \cos\biggl(
      {{1}\over{2}}t \sqrt{\mit\Gamma^2+\mit\Delta^2}
      \biggr)
 \sin\biggl(
        {{1}\over{2}}\omega t
       \biggr) ,
\cr
\displaystyle 
 s_0(t) = {{\mit\Delta}\over{\sqrt{\mit\Gamma^2+\mit\Delta^2}}}
 \sin\biggl(
      {{1}\over{2}}t \sqrt{\mit\Gamma^2+\mit\Delta^2}
      \biggr)
 \sin\biggl(
        {{1}\over{2}}\omega t
       \biggr)  
\cr
\displaystyle
 +
  \cos\biggl(
      {{1}\over{2}}t \sqrt{\mit\Gamma^2+\mit\Delta^2}
      \biggr)
 \cos\biggl(
        {{1}\over{2}}\omega t
       \biggr) ,
         }
\right.       
\label{eq=39}
\end{equation}
 Here the parameters
\begin{equation}
  {\mit\Delta}     = \omega - {\mit\Omega}_*\cos(\theta),~~~~
  {\mit\Gamma} = {\mit\Omega}_*\sin(\theta)
\label{eq=40}
\end{equation} 
 specify the position and the width of a resonance maximum, where
\begin{equation}
{\mit\Omega}_* = \gamma_n \sqrt{B_z^2 + b^2},~~~~
\theta = \arctan(b/B_z)
\label{eq=41}
\end{equation} 
 are the Larmor precession frequency and the apex angle of the cone described by the vector $\vec B$, respectively~\cite{IoffeEtAl1991}.
 
 The polarization vector for the neutron passing through this helical magnetic configuration reads
\begin{equation}
  {\vec P}(t) = R(s_0(t), s_x(t), s_y(t),\pm s_z(t)){\vec P}(0).
\label{eq=42}
\end{equation} 
Right now we can obtain solutions  in the case of the resonance ${\mit\Delta}=0$  for both signs, $\pm$, at the component $s_z(t)$:
\begin{eqnarray}
\label{eq=43}
 {\rm sign~(+)~at~}s_z(t)&:&
 \left\{
    \matrix{
        P_x(t) = -\sin({\mit\Gamma}t)\sin(\omega t), \cr
        P_y(t) = \sin({\mit\Gamma}t)\cos(\omega t),\hskip10pt \cr
        P_z(t) = \cos({\mit\Gamma}t), \hskip42pt
               }
 \right. 
\\ 
 {\rm sign~(-)~at~}s_z(t)&:&
 \left\{
    \matrix{
        P_x(t) = 0, \hskip34pt \cr
        P_y(t) = -\sin({\mit\Gamma}t), \cr
        P_z(t) = \cos({\mit\Gamma}t).\hskip10pt
               }
 \right.  
\label{eq=44}
\end{eqnarray}
These solutions are shown in Fig.~\ref{fig=5}. The solution~(\ref{eq=43}) draws the helical ring colored in red, and the solution~(\ref{eq=44}) gives the simple circle colored in blue.

\begin{figure}[htb!]
  \begin{picture}(180,80)(0,5)
      \includegraphics[scale=0.5]{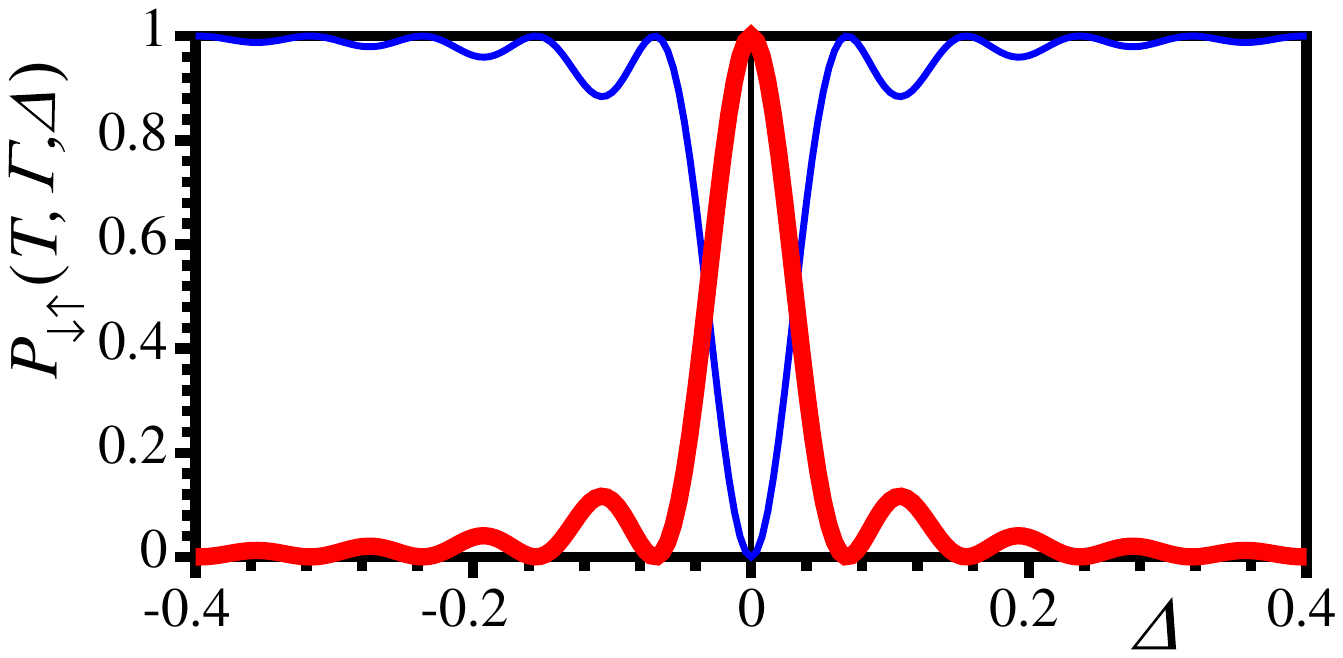}
  \end{picture}
  \caption{
   Spin-flip resonance curves $P_{\downarrow}(t,{\mit\Gamma},{\mit\Delta})$  (colored in red) and 
   $P_{\uparrow}(t,{\mit\Gamma},{\mit\Delta})$  (colored in blue) versus ${\mit\Delta}$. Here ${\mit\Gamma}=0.04$ and $T=\pi/{\mit\Gamma}$.
  }
  \label{fig=6}
\end{figure}
The spin-flip resonance curve (turn spin down probability) for neutrons passing through PMS~(\ref{eq=38}) has a standard form~\cite{AgamalyanEtAl1988}:
\begin{equation}
   P_{\downarrow}(T,{\mit\Gamma},{\mit\Delta}) = |\varphi_{\downarrow}|^2 =   u_x^2(T) + u_y^2(T)
  = {{\mit\Gamma^2}\over{\mit\Gamma^2+\mit\Delta^2}}
   \sin^2\biggl(
      {{T}\over{2}}\sqrt{\displaystyle{\mit\Gamma^2+\mit\Delta^2}}
            \biggr).
\label{eq=45}
\end{equation} 
This curve  colored in red is shown in Fig.~{\ref{fig=6}} at the parameters ${\mit\Gamma} = 0.04$, ${\mit\Delta}$ ranging from $-0.4$ to $0.4$, and  $T=\pi/{\mit\Gamma}$ is the passing time through the periodic magnetic structure corresponding to the complete resonance. 
Also one can compute the revolution spin up probability, $P_{\uparrow}(T,{\mit\Gamma},{\mit\Delta})$. This probability shown in Fig.~{\ref{fig=6}} in blue has the following view
\begin{eqnarray}
\nonumber
   P_{\uparrow}(T,{\mit\Gamma},{\mit\Delta}) &=& |\varphi_{\uparrow}|^2 =   u_z^2(T) + u_0^2(T)
\\
&=&   {{\mit\Gamma^2}\over{\mit\Gamma^2+\mit\Delta^2}}
   \cos^2\biggl(
      {{T}\over{2}}\sqrt{\displaystyle{\mit\Gamma^2+\mit\Delta^2}}
            \biggr)
 +    {{\mit\Delta^2}\over{\mit\Gamma^2+\mit\Delta^2}}.
\label{eq=46}
\end{eqnarray}
 One can see that sum of these probabilities, $P_{\downarrow}$ and $P_{\uparrow}$, is 1.

Thus we have described the rotation of spin--1/2 as motion of the quaternion vector on the sphere ${\mathcal S}^3$. It turns, that motion of this vector on the sphere ${\mathcal S}^3$ is equivalent to rotation on the sphere ${\mathcal S}^2$ of the polarization vector  loaded by the flag $u_0(t)$. Let us note further that this mathematical object, called spin-1/2, is coated by self-organized vortex flows of superfluid quantum particles, the Bose particles, which come from the solution of the vorticity equation~(\ref{eq=11}). This remark hints that the bare magnetic moment  of the particle with spin 1/2 (the Bohr magneton, $\mu_{B}\approx -9.27401452557\times10^{-24}$ J$\cdot$T$^{-1}$ for electron, for example) should be corrected by extra terms that lead to the anomalous magnetic moment $\mu_{e}\approx -9.28476377\times10^{-24}$ J$\cdot$T$^{-1}$~\cite{Sbitnev2016c}.

As for the continuity equation we should put into consideration a pair of equations. Together they describe the continuity of flows in 6D space ${\mathcal R}^{3}\times{\mathcal S}^3$ and look as follows
\begin{eqnarray}
\label{eq=47}
&&    {{\partial}\over{\partial\,t}}\rho({\vec r},t) + (\nabla\rho({\vec r},t)){\vec v} = 0,  \\
&&  {{d}\over{d\,t}}|\varphi(t)\rangle = -{{\mu_{e}}\over{\hbar}}({\overrightarrow\eta}\cdot{\vec B}) |\varphi(t)\rangle,
\label{eq=48}
\end{eqnarray}
 where $({\overrightarrow\eta}\cdot{\vec B})$ and $|\varphi(t)\rangle$ are given by~(\ref{eq=36}) and~(\ref{eq=37}).

\section{\label{sec4}Schr\"{o}dinger-Pauli equation}

Let us return to Eq.~(\ref{eq=10}) and note that the rightmost term in this equation contains the velocity 
${\vec{\mathit v}}={\vec{\mathit v}}_{S}+{\vec{\mathit v}}_{R}$. By applying the curl operator we obtain ${\vec\omega}=[\nabla\times{\vec{\mathit v}}_{R}]$, but $[\nabla\times{\vec{\mathit v}}_{S}]$ vanishes. On the other hand, by applying the divergence operator we get  $(\nabla\!\cdot\!{\vec{\mathit v}}_{R})=0$ and $(\nabla\!\cdot\!{\vec{\mathit v}}_{S})$ does not vanish. As a result, we can rewrite Eq.~(\ref{eq=10}) in the following view
\begin{equation}
      {\frac{\partial\;}{\partial\,t}} S + {\frac{1}{2m_{}}} (\nabla S - e{\vec A})^2 + e\phi
     + (U + Q)
     - {\nu}(t)m(\nabla {\vec{\mathit v}})
  = C_0.
\label{eq=49}
\end{equation}
Here $C_0=\omega_0\hbar$ is an integration constant that adds an arbitrary phase $\phi=\omega_0t$ to a wave function $\Psi$, see below the \Schrodinger equation. It is interesting to note that de Broglie identified the frequency $\omega_0$ as a clock frequency of a periodic or pulsating phenomenon that extends throughout all space~\cite{Osche2011}.

Accurate to two extra terms, $Q$ and $\nu(t)m\nabla{\vec{\mathit v}}$, this equation represents the classical Hamilton-Jacobi equation. When adding the quantum potential $Q$ it turns into the quantum Hamilton-Jacobi equation.  The term $\nu(t)m\nabla{\vec{\mathit v}}$ does not influence the following computation and we shall not discuss it yet.
Observe that the quantum Hamilton-Jacobi equation~(\ref{eq=49}) together with the continuity equations~(\ref{eq=47}) and~(\ref{eq=48}) are extracted from the following  Schr\"{o}\-dinger-Pauli-like equation
\begin{eqnarray}
 \nonumber
 &&
 \underbrace{
      {\bf i}\hbar{{\partial}\over{\partial\,t}}|\Psi({\vec r},t)\rangle
   = \Biggl(
           {{1}\over{2m}}
           \biggl(
              -{\bf i}\hbar\nabla - e{\vec A}
           \biggr)^{2}
           +e\phi   +  U({\vec r})  + \overbrace{\nu(t)mf(\rho)}^{(a)}
      \Biggr)
      |\Psi({\vec r},t)\rangle
                   }_{\rm Schr{o}dinger-like~equation}
\\
  &-& \underbrace{
       {\bf i}\mu_{e} ({\overrightarrow\eta}\!\cdot\!{\vec B}({\vec r},t)) |\Psi({\vec r},t)\rangle
                  }_{\rm Stern-Gerlach~term}
  \;+\; C_0|\Psi({\vec r},t)\rangle                   
\label{eq=50}
\end{eqnarray}
 when substituting in this equation the wave function represented in the following polar form
\begin{equation}
    |\Psi({\vec r},t)\rangle = \sqrt{\rho({\vec r},t)}|\varphi(t)\rangle\!\cdot\!\exp\{ {\bf i}S({\vec r},t)/\hbar \}.
\label{eq=51}
\end{equation} 
After substituting this wave function into Eq.~(\ref{eq=50}) and separating solutions on imaginary and real parts we come to Eqs.~(\ref{eq=47}),~(\ref{eq=48}) and~(\ref{eq=49})~\cite{Sbitnev2008}.

Distinction from the original Schr\"{o}dinger-Pauli equation is due to presence of the extra term $\nu(t)mf(\rho)$ embraced by the  bracket~(a) in Eq.~(\ref{eq=50}).  The term $f(\rho)$ comes from the continuity equation~\cite{Sbitnev2016b}
\begin{equation}
 f(\rho) = {{d}\over{d\,t}}\ln(\rho) =
 {{\partial}\over{\partial\,t}}\ln(\rho) + {\mathit{\vec v}}\nabla\ln(\rho) = -\nabla{\mathit{\vec v}}.
\label{eq=52}
\end{equation}
 Observe that the term
\begin{equation}
  \nu(t)m f(\rho)|\Psi\rangle = \nu(t)m f(|\Psi|^2)|\Psi\rangle
\label{eq=53}  
\end{equation} 
  in Eq.~(\ref{eq=50}) makes the Schr\"{o}dinger equation similar to 
  the  Gross-Pitaevskii equation~\cite{AbidEtAl2003,RobertsBerloff2001}.
   Here the function $f(\rho)=d\ln(\rho)/dt$ expanded into Tailor series as a polynomial  of $\rho=|\Psi|^2$ with real coefficients gives possibility to use approximation methods~\cite{AbidEtAl2003} in analyzing the behavior of Bose-Einstein  condensates
existing in the superfluid quantum space.
   
Since $\nu(t)$ is a fluctuating function equal to zero when averaging on time with non-zero variance~(\ref{eq=2}), it can mean that the term $\nu(t)mf(|\Psi|^2)$ describes energetic exchange between the particle and the zero-point vacuum energy fluctuations. One can conclude that the \Schrodinger-Pauli equation loaded by the non-linear term $\nu(t)mf(|\Psi|^2)|\Psi\rangle$ represents the Langevin equation with a source of color-noise.
 
The magnetic term $ ({\overrightarrow\eta}\!\cdot\!{\vec B}({\vec r},t))$ in Eq.~(\ref{eq=50}) represents itself the real $4\!\times\!4$ matrix~(\ref{eq=36}) and $|\varphi(t)\rangle$ is the real 4-vector~(\ref{eq=37}). As the transition to the SU(2) group spinors occurs, one comes to the familiar \Schrodinger-Pauli equation where the imaginary factor ${\bf i}$ at the magnetic term is absent and instead of the quaternion matrices~(\ref{eq=24}) the Pauli matrices~(\ref{eq=17}) are presented.
  
An unusual representation of the Stern-Gerlach term through the quaternion matrices where the magnetic field looks as a $4\times4$ ordered matrix~(\ref{eq=36}) with real coefficients invites to consider for completeness the Maxwell's electromagnetic field theory  in the quaternion algebra.

\section{\label{sec5}Maxwell's electromagnetic field theory  in the quaternion basis}   

First, let us define differential operators having the following quaternion representation
\begin{equation}
\left.
   \matrix{
        {\mathcal D}\hskip6pt &\, ={\bf i}c^{-1}\partial_t\eta_0
              \:+ \partial_x\eta_x\: + \partial_y\eta_y\: + \partial_z\eta_z, \cr
        {\mathcal D}^{\bf T}\!& ={\bf i}c^{-1}\partial_t\eta_0^{\bf T}
              + \partial_x\eta_x^{\bf T} + \partial_y\eta_y^{\bf T} + \partial_z\eta_z^{\bf T} \cr
          &\, =   {\bf i}c^{-1}\partial_t\eta_0\:
              - \partial_x\eta_x\: - \partial_y\eta_y\: - \partial_z\eta_z.
              }
\right.
\label{eq=54}
\end{equation}
 Here $\partial_t =\partial/\partial t$, $\partial_x =\partial/\partial x$, etc.,  $c$ is the speed of light, and sign ${\bf T}$ means the transposition. The d'Alembertian or the wave operator with the negative metric signature $\{-1,+1,+1,+1\}$ ($g_{00}=-1$, $g_{11}=g_{22}=g_{33}=1$) reads
\begin{equation}
   {\mathcal D}^{\bf T} {\mathcal D} = \
    (-c^{-2}\partial_t^2 + \partial_x^2 + \partial_y^2 + \partial_z^2)\eta_0
\label{eq=55}
\end{equation} 
 An electromagnetic four-potential looks as follows
\begin{equation}
  {\vec\Phi} = {\bf i}\phi\eta_0 + A_x\eta_x + A_y\eta_y + A_z\eta_z.
\label{eq=56}  
\end{equation} 
 The electromagnetic potential underlies the definition of the observable magnetic and electric fields
\begin{equation}
\left.
   \matrix{
       {\vec B} &\hskip-34pt = [\nabla\times{\vec A}],  \cr\cr
       {\vec E} &\hskip-6pt   = -\nabla\phi - {\displaystyle {{1}\over{c}}{{\partial {\vec A}}\over{\partial\,t}}}.
             }
\right.
\label{eq=57}
\end{equation} 
  These formulas are given in Gaussian units. Observe that these fields are invariant with respect to the gauge transformation $\phi\rightarrow\phi-c^{-1}\partial_t \psi$ and ${\vec A}\rightarrow{\vec A}+ \nabla\psi$, (here $\psi$ is an arbitrary scalar field).
  In order to avoid the ambiguousness in representing the electromagnetic potential, the Lorentz gauge transformation is introduced 
\begin{equation}
 -{{1}\over{4}}{\rm trace}\;{\mathcal D}{\vec\Phi} ={{1}\over{c}}\partial_t\phi
 + \partial_xA_x  + \partial_yA_y  + \partial_zA_z = 0. 
\label{eq=58}
\end{equation}  
Multiplications of the matrices $\eta_x$, $\eta_y$, $\eta_z$ are shown in Eq.~(\ref{eq=25}).

Let us now write out  the product ${\mathcal D}\!\cdot\!{\vec\Phi}$ in details
\begin{eqnarray}
\nonumber
 {\mathcal D}\!\cdot\!{\vec\Phi} &=& 
 \biggl(
   -{{1}\over{c}}{{\partial\phi}\over{\partial t}}
   - {{\partial A_x}\over{\partial x}}
   - {{\partial A_y}\over{\partial y}}
   - {{\partial A_z}\over{\partial z}}
 \biggr)\eta_0
\\ \nonumber
 &+&
 \biggl(
   {\bf i}
   \biggl(
     {{\partial\phi}\over{\partial x}}+{{\partial A_x}\over{c\partial t}}
   \biggr)
- \biggl(
   {{\partial A_z}\over{\partial y}} - {{\partial A_y}\over{\partial z}}
  \biggr)
 \biggr)\eta_x 
\\ \nonumber
 &+&
 \biggl(
   {\bf i}
   \biggl(
     {{\partial\phi}\over{\partial y}}+{{\partial A_y}\over{c\partial t}}
   \biggr)
- \biggl(
   {{\partial A_x}\over{\partial z}} - {{\partial A_z}\over{\partial x}}
  \biggr)
 \biggr)\eta_y
\\ 
 &+&
 \biggl(
   {\bf i}
   \biggl(
     {{\partial\phi}\over{\partial z}}+{{\partial A_z}\over{c\partial t}}
   \biggr)
- \biggl(
   {{\partial A_y}\over{\partial x}} - {{\partial A_x}\over{\partial y}}
  \biggr)
 \biggr)\eta_z 
\label{eq=59}
\end{eqnarray}
One can see that the expression at the quaternion $\eta_0$ represents the Lorenz gauge when it is zero. There remain only the expressions at the quaternions $\eta_x$, $\eta_y$, $\eta_z$. According to the equations given in~(\ref{eq=57}), these expressions represent $x$, $y$, $z$ components of the electric and magnetic fields. So, the above equation represents the electromagnetic tensor
\begin{equation}
 {\vec F}_{\rm EM} = {\mathcal D}\!\cdot\!{\vec\Phi} =
\left(
  \matrix{
                       0     & ~~B_x - {\bf i}E_x  & ~~B_y - {\bf i}E_y   & ~~B_z - {\bf i}E_z \cr
    -B_x + {\bf i}E_x  & 0                           &    -B_z + {\bf i}E_z  & ~~B_y - {\bf i}E_y \cr
    -B_y + {\bf i}E_y  &~~B_z -{\bf i}E_z    &      0                       &    -B_x +{\bf i}E_x \cr
    -B_z + {\bf i}E_z  &   -B_y+{\bf i}E_y     & ~~B_x - {\bf i}E_x   &     0                     \cr
            }
\right) .
\label{eq=60}
\end{equation}

\subsection{\label{sec51}Electromagnetic fields} 

The whole set of Maxwell's equations consists of two pairs of equations. The first pair is represented by Gauss's law for magnetism and  Faraday's law:
\begin{equation}
\left.
   \matrix{
        (\nabla\!\cdot\!{\vec B}) = 0, \cr \cr
        {\displaystyle
           [\nabla\times{\vec E}] + {{1}\over{c}}{{\partial}\over{\partial t}}{\vec B}} = 0.
              }
\right.
\label{eq=61}
\end{equation}
The second pair contains Maxwell-Ampere and Gauss's laws:
\begin{equation}
\left.
   \matrix{
        {\displaystyle
           [\nabla\times{\vec B}] - {{1}\over{c}}{{\partial}\over{\partial t}}{\vec E} = {{4\pi}\over{c}}{\vec j},  }   \cr \cr
        (\nabla\!\cdot\!{\vec E}) = 4\pi\rho.
            }
\right.
\label{eq=62}
\end{equation}
 The second pair contains extra terms - the charge density
\begin{equation}
  \rho = e\,\delta({\vec r}-{\vec r}_0)
\label{eq=63}
\end{equation} 
and the 3D current density
\begin{equation}
 {\vec j} = \rho{\mathit{\vec v}}.
\label{eq=64}
\end{equation}
 Here $e$ is the charge and ${\mathit{\vec v}}$ is the charge velocity in the vicinity of the point ${\vec r}$.
 Motion of the charge submits to the continuity equation
\begin{equation}
   {{\partial\rho}\over{\partial t}} + (\nabla\!\cdot\!{\vec j}) = 0.
\label{eq=65}
\end{equation} 

Let us define the 4D current density as follows
\begin{equation}
 {\vec J} = -{\bf i}c\rho\eta_0 + j_x\eta_x + j_y\eta_y + j_z\eta_z.
\label{eq=66}
\end{equation}
 The continuity equation in this case takes the following view
\begin{equation}
 {{1}\over{4}}{\rm trace}\;{\mathcal D}^{\bf T}\!\cdot\!{\vec J} =
 {\partial_t\rho} + {\partial_x j_x} + {\partial_y j_y} + {\partial_z j_z} = 0.
\label{eq=67}
\end{equation} 
By computing the product ${\mathcal D}\!\cdot\!{\vec F}_{EM}$
\begin{eqnarray}
\nonumber
&& {\mathcal D}\!\cdot\!{\vec F}_{EM} =
\{       ({\partial_x B_x} + {\partial_x B_x} + {\partial_x B_x})
-{\bf i}({\partial_x E_x} + {\partial_x E_x} + {\partial_x E_x})
\}\eta_0  
\\ \nonumber &&+ 
\biggl\{\!\!
  \biggl(
    -{{1}\over{c}}{\partial_t E_x} + ({\partial_y B_z} - {\partial_z B_y})
  \biggr)
 +{\bf i}  \biggl(
    -{{1}\over{c}}{\partial_t B_x} - ({\partial_y E_z} - {\partial_z E_y})
  \biggr)
\!\!\biggr\}\eta_x
\\ \nonumber &&+ 
\biggl\{\!\!
  \biggl(
    -{{1}\over{c}}{\partial_t E_y} + ({\partial_z B_x} - {\partial_x B_z})
  \biggr)
 +{\bf i}  \biggl(
    -{{1}\over{c}}{\partial_t B_y} - ({\partial_z E_x} - {\partial_x E_z})
  \biggr)
\!\!\biggr\}\eta_y
\\ \nonumber &&+ 
\biggl\{\!\!
  \biggl(
    -{{1}\over{c}}{\partial_t E_z} + ({\partial_x B_y} - {\partial_y B_x})
  \biggr)
 +{\bf i}  \biggl(
    -{{1}\over{c}}{\partial_t B_z} - ({\partial_x E_y} - {\partial_y E_x})
  \biggr)
\!\!\biggr\}\eta_z
\\ &&\hskip22pt =
{{4\pi}\over{c}}(-{\bf i}c\rho\eta_0 + j_x\eta_x + j_y\eta_y + j_z\eta_z)
\label{eq=68}
\end{eqnarray}
we find that both pairs of equations, Eqs.~(\ref{eq=61}) and~(\ref{eq=62}), result from
\begin{equation}
 {\mathcal D}\!\cdot\!{\vec F}_{EM} = {{4\pi}\over{c}}{\vec J}.
\label{eq=69}
\end{equation}
 By applying the operator ${\mathcal D}^{\bf T}$ from the left to this equation we get the electromagnetic wave equation:
\begin{equation}
  {\mathcal D}^{\bf T} {\mathcal D}\!\cdot\!{\vec F}_{EM} = 0 \hskip24pt\Rightarrow\hskip24pt
  \biggl(
    \nabla^2 - {{1}\over{c^2}}{{\partial^2}\over{\partial t^2}}
  \biggr)({\vec B} - {\bf i}{\vec E}) = 0.
\label{eq=70}
\end{equation} 
 In fact, here we have got two separate wave equations, one for the magnetic field the other for the electric field.
 
\subsection{\label{sec52}Lorentz transformation} 

The Lorentz transformation is that of coordinates between two  frames of reference that move at constant velocity relative to each other. The Lorentz transformation of the 4-dimensional space-time spanned on the quaternion basis $(\eta_0, \eta_x, \eta_y, \eta_z)$  is carried out by matrices of the view
\begin{equation}
\left.
  \matrix{
       L~~             = \nu_0\eta_0 + \nu_x\eta_x + \nu_y\eta_y + \nu_z\eta_z, \cr
       L^{\bf T} = \nu_0\eta_0 - \nu_x\eta_x  - \nu_y\eta_y  - \nu_z\eta_z
            }
\right.
\label{eq=71}
\end{equation}
 We require that there is always the equality $L\!\cdot\!L^{\bf T}=L^{\bf T}\!\cdot\!L=\eta_0$. From here it follows that the parameters $\nu_0$, $\nu_x$, $\nu_y$, $\nu_z$ submit to the following constraint
\begin{equation}
  \nu_0^2 + \nu_x^2 + \nu_y^2 + \nu_z^2 = 1.
\label{eq=72}
\end{equation} 
Two types transformations can be written. They are rotation of the coordinate system in three-dimensional space and the transition to the moving coordinate system (boost)~\cite{PenroseRindler1984,PenroseRindler1986}:
\begin{itemize}
  \item[$\bullet$] transformations of the coordinate system in the Minkowski space associated with rotation about the axis 
  ${\vec m}=(m_x,m_y,m_z)$ on an angle $\alpha$ are realized by the following trigonometric functions
  \begin{equation}
     \nu_0 = \cos\biggl(
                          {{\alpha}\over{2}}
                        \biggr), \hskip44pt
    {\vec\nu}= {\vec m} \cdot                        
                  \sin\biggl(
                          {{\alpha}\over{2}}
                        \biggr);
  \label{eq=73}
  \end{equation}
 \item[$\bullet$] transformations of the coordinate system in the Minkowski space associated with the shift in the direction of the axis  ${\vec m}=(m_x,m_y,m_z)$ on a boost $\beta={\mathit v}/c$ (${\mathit v}$ is a speed of the moving laboratory coordinate system and $c$ is the speed of light) are realized by the following hyperbolic functions
 \begin{equation}
     \nu_0 = \cosh\biggl(
                          {{\beta}\over{2}}
                        \biggr), \hskip44pt
    {\vec\nu}= {\vec m} \cdot                        
                  \sinh\biggl(
                          {{\beta}\over{2}}
                        \biggr);
 \label{eq=74}
 \end{equation}
\end{itemize}

During the transition to the new coordinate system the electromagnetic field tensor ${\vec F}_{EM}$ is subjected to the transformation
\begin{equation}
{\vec F}_{EM}^{\prime} = L\!\cdot\!{\vec F}_{EM}\!\cdot\!L^{\bf T}.
\label{eq=75}
\end{equation}
After series of computations we come to the following transformation of the complex electromagnetic field 
${\vec F}=-{\vec B}+{\bf i}{\vec E}$:
\begin{equation}
  {\vec F}^{\prime} =
  \left(
     \matrix{
        -B_x^{\prime} + {\bf i}E_x^{\prime} \cr
        -B_y^{\prime} + {\bf i}E_y^{\prime} \cr
        -B_z^{\prime} + {\bf i}E_z^{\prime} 
               }
  \right) =
  R({\vec\nu})\!
  \left(
     \matrix{
        -B_x + {\bf i}E_x \cr
        -B_y + {\bf i}E_y \cr
        -B_z + {\bf i}E_z 
               }
  \right) =
   R({\vec\nu}){\vec F}.
\label{eq=76}
\end{equation}
 The $3\!\times\!3$ matrix $R({\vec\nu})= R(\nu_0,\nu_x,\nu_y,\nu_z)$ resulting from the computations of Eq.~(\ref{eq=75}) has the following view:
\begin{eqnarray}
\nonumber
&& R(\nu_0,\nu_x,\nu_y,\nu_z) = 2\nu_0
   \left(
      \matrix{
            ~0    & ~~\nu_z  &     -\nu_y \cr
            -\nu_z  &   ~0     &  ~~\nu_x \cr
         ~~\nu_y  &    -\nu_x  &    ~0    \cr
                 } 
   \right)
\\ 
   && + 
  \left(
     \matrix{
     1-2(\nu_y^2+\nu_z^2) & 2\nu_x\nu_y                & 2\nu_z\nu_x                  \cr
     2\nu_y\nu_x                & 1-2(\nu_z^2+\nu_x^2) & 2\nu_y\nu_z                  \cr
     2\nu_z\nu_x                & 2\nu_x\nu_y                 & 1-2(\nu_x^2+\nu_y^2)  \cr
               }
  \right) .
\label{eq=77}  
\end{eqnarray}
By writing out in clear view the transformation~(\ref{eq=76}) we get the following forms
\begin{itemize}
\item[$\bullet$] 
 when rotating the coordinate system about the axis ${\vec m}$ on the angle $\alpha$
\begin{equation}
   {\vec F}^{\prime} = {\vec F}\cdot\cos(\alpha) + {\vec m}({\vec m}\cdot{\vec F})\cdot(1-\cos(\alpha))
   - [{\vec m}\times{\vec F}]\cdot\sin(\alpha).
\label{eq=78}
\end{equation} 
One can see that the electric and magnetic fields undergo rotations independently from each other;
\item[$\bullet$]
when boosting into the coordinate system  moving with the velocity ${\mathit{\vec v}}$ along the direction
 ${\vec m}={\mathit{\vec v}}/|{\mathit v}|$ with respect to the initial coordinate system,  the Lorentz transformation shifts by a value $\phi$  (it is a hyperbolic angle), to be named the rapidity  
\begin{equation}
   {\vec F}^{\prime} = {\vec F}\cdot\cosh(\phi) + {\vec m}({\vec m}\cdot{\vec F})\cdot(1-\cosh(\phi))
   - {\bf i}[{\vec m}\times{\vec F}]\cdot\sinh(\phi).
\label{eq=79}
\end{equation} 
Note that $\cosh(\phi)=\gamma = 1/\sqrt{1-\beta^2}$ is the Lorentz factor and $\sinh(\phi)=\beta\gamma$, 
here $\beta={\mathit v}/c$ is the velocity coefficient into the ${\vec m}$-direction.
By expressing this formula through the electric and magnetic fields  we get
\begin{eqnarray}
\label{eq=80}
 && 
 {\vec E}^{\prime} = \gamma{\vec E} - (\gamma-1)({\vec E}\!\cdot\!{\mathit{\vec v}}){\mathit{\vec v}}/{\mathit v}^2
                            + (\gamma/c)[{\mathit{\vec v}}\times{\vec B}],
 \\ 
 &&
 {\vec B}^{\prime} = \gamma{\vec B} - (\gamma-1)({\vec B}\!\cdot\!{\mathit{\vec v}}){\mathit{\vec v}}/{\mathit v}^2
                             - (\gamma/c)[{\mathit{\vec v}}\times{\vec E}].
\label{eq=81}
\end{eqnarray}
\end{itemize}

\subsection{\label{sec53}Quadratic forms of the electromagnetic field tensor}

Two quadratic forms can be composed from the electromagnetic tensor. 

The first quadratic form
\begin{equation}
{{1}\over{2}} {\vec F}_{EM}{\vec F}_{EM}^{\dagger} =
 -W_0\eta_0 + {\bf i}W_x\eta_x + {\bf i}W_y\eta_y + {\bf i}W_z\eta_z
\label{eq=82}
\end{equation}
contains information about the density and flux of the electromagnetic energy.  Here $\dagger$ is the sign of complex conjugation, $W_0 = (E^2+B^2)/2$ is the energy density, and ${\vec W}=[{\vec E}\times{\vec B}]$ describes the energy flux in the direction perpendicular to the fields ${\vec E}$ and ${\vec B}$. Accurate to the divider $\mu_0$ (the vacuum permeability) the vector ${\vec W}$ represents the Poynting vector.

The second quadratic form  reads
\begin{equation}
{{1}\over{2}} {\vec F}_{EM}^{\bf T}{\vec F}_{EM} =
 {{(B^2-E^2)}\over{2}}\eta_0 - {\bf i}({\vec E}\!\cdot\!{\vec B})\eta_0.
\label{eq=83}
\end{equation}
It gives two invariants with respect to the Lorentz transformations.
The first invariant $I_1=(B^2-E^2)/2$ is the scalar and the second invariant $I_2 = ({\vec E}\!\cdot\!{\vec B})$ is the pseudoscalar.
 
 The electromagnetic tensor~(\ref{eq=60}) looks in unusual manner. The magnetic field represents a real part of the tensor, while the electric field appears as an imaginary part. At that,  the both fields are represented by a symmetric way: ${\vec B}\leftrightarrow-{\bf i}{\vec E}$. Such an organization of the electromagnetic tensor is due to different roles of the magnetic and electric fields played by them at moving magnetic and electric dipoles in the superfluid quantum space. The magnetic dipoles undergo rotations in 3D space under influence of the magnetic field. While the electric dipoles experience alignments along force lines of the electric field with attraction of the charges by opposite poles. This alignment acts like the boosting through the hyperbolic accelerating shifts. That is, when  oppositely charged dipole arms are moving with acceleration until they align along the force lines.
 
\section{\label{sec6}Conclusion}   
   
The modified Navier-Stokes equation describing the  velocity field of particles in the superfluid quantum space and the continuity equation describing the conservation of the probability density distribution of the particles moving through this 3D space are those underlaying the \Schrodinger equation. For getting the \Schrodinger-Pauli equation we need to take into account the motion of the particle spin in a magnetic field. It leads to expanding the continuity equation acting in 3D Euclidean space up to the continuity equation acting in 6-dimensional space. This space is the product of the real 3D Euclidean space by the 3D sphere of unit radius, namely, ${\mathcal R}^3{\otimes}{\mathcal S}^3$. The latter space spanned by quaternion basis of 4x4 matrices is the space in which rotations of spin are described in detail. 

 We know that rotations of a spin-1/2 particle are described by SU(2) group, the special unitary group of $2\times2$ unimodular matrices. This group is isomorphic to the group of quaternions with norm 1. Its transformations are diffeomorphic to motions on the 3D sphere. These motions can be represented on the 2D sphere if we assume that the tip of the 3D vector is equipped by a flag. Orientation of the flag is extra dimension to the vector, $u_0=\cos(\theta/2)$, having three native axes $(x, y, z)$. In the whole all four degrees of the freedom are linked by the normalization condition on the unit, $u_0^2+u_x^2+u_y^2+u_z^2=1$. 
As an example, we have considered the spin resonance of neutrons passing  through the spatially periodic magnetic fields. The spin-1/2 at the resonance should have the double passing the spin tip through the top pole of the 2D sphere in order to complete the whole revolution of the spin flag. At the first revolution its orientation gets the angle $\pi$, and after the second revolution it reaches the full turn equal to $2\pi$.   At the spin resonance the spin tip on the ${\mathcal S}^2$ sphere draws a perfect helical loop 
described in~\cite{Sbitnev2017}.

Note that in a physical experiment we can measure only two spin states - either spin-up or spin-down with respect to orientation of the magnetic field.
Therefore the space ${\mathcal R}^{3}\otimes{\mathcal S}^{3}$ splits into two foliations 
${\mathcal R}^{3}_{\uparrow}$ and ${\mathcal R}^{3}_{\downarrow}$.
In the first case the energy of the particle is $E+\mu_{e}B_{z}$ in the second case it is $E-\mu_{e}B_{z}$, where $E=m{\mathit v}^2/2$ is the kinetic energy of the particle and $\pm\mu_{e}B_{z}$ represent two levels of the splitting, conditionally, the levels up and down.
Here the $z$-axis is adopted as the axis of quantization. At such a measurement we cannot distinguish the orientation of the spin flag.
We loose many information about motion of the spin on the ${\mathcal S}^3$ sphere. However, these observations can be reached in the interference experiment.

In the same quaternion basis Maxwell's theory of electromagnetism looks by the very symmetric way. Both magnetic and electric fields manifest themselves as a single complex field ${\vec B}+{\bf i}{\vec E}$ having the very symmetric electromagnetic tensor.  
Both fields can transform to each other  under the Lorentz transformations.

The real part of the tensor is the magnetic field completing the $4\!\times\!4$ matrix. Quaternion organization of the latter determines the interaction of the spin-1/2 particle with the magnetic field. In the quaternion basis we get an elegant description of the spin rotation in the magnetic field. On the other hand, the imaginary part of the tensor gives the electric field completing the same $4\!\times\!4$ matrix. This matrix determines the interaction of an electric dipole with the electric field. Motion of the electric dipole under influence of the electric field has an accelerated character until the dipole aligns along the electric filed.

\begin{acknowledgements}
The author thanks Marco Fedi for useful and valuable remarks and offers.
\end{acknowledgements}


%
%

\end{document}